\documentclass{article}
\usepackage{ijcai21}
\usepackage{time}
\usepackage{times}
\usepackage{soul}
\usepackage{url}
\usepackage[hidelinks]{hyperref}
\usepackage[utf8]{inputenc}
\usepackage[small]{caption}
\usepackage{graphicx}
\usepackage{subfigure}
\usepackage{float}
\usepackage{amsmath}
\usepackage{amsthm}
\usepackage{booktabs}
\usepackage{algorithm}
\usepackage{algorithmic}
\usepackage{enumerate}
\usepackage{booktabs}
\usepackage{multirow} 
\usepackage{stfloats}
\usepackage{fancyvrb,cprotect}
\usepackage{acronym} %for ac command
\usepackage{amssymb}
\usepackage{bm}

\acrodef{GCN}{Graph Convolution Network}
\acrodef{GCNs}{Graph Convolution Networks}
\acrodef{SCRM}{Self-attention-based Cross-domain Recommendation Machine}
\acrodef{SCSR}{Shared-account Cross-domain Sequential Recommendation}
\acrodef{SR}{Sequential Recommendation}
\acrodef{RNNs}{Recurrent Neural Networks}
\acrodef{RNN}{Recurrent neural network}
\acrodef{CNN}{Convolutional neural network}
\acrodef{MRR}{Mean Reciprocal Rank}
\acrodef{CF}{Collaborative Filtering}
\acrodef{SA}{Self-Attention}
\acrodef{CR}{Cross-domain Recommendation}
\acrodef{DA-GCN}{Domain-Aware Graph Convolutional Network}
\acrodef{CDS}{Cross-Domain Sequence}
\acrodef{CSR}{Cross-domain Sequential Recommendation}
\acrodef{SA}{Shared-Account}

\title{DA-GCN: A Domain-aware Attentive Graph Convolution Network for Shared-account Cross-domain Sequential Recommendation}

\author{
Lei Guo$^1$
\and
Li Tang$^1$\and
Tong Chen$^{2}$\and
Lei Zhu$^1$ \and 
Quoc Viet Hung Nguyen$^3$ \and
Hongzhi Yin$^{2}$\footnote{Corresponding author}
\affiliations
$^1$Shandong Normal University, China\\
$^2$The University of Queensland, Australia\\
$^3$Griffith University, Australia
\emails
leiguo.cs@gmail.com, litang96@126.com, tong.chen@uq.edu.au,
\{leizhu0608, quocviethung1\}@gmail.com,
h.yin1@uq.edu.au
}
\begin{document}

\maketitle

\begin{abstract}

\ac{SCSR} is the task of recommending the next item based on a sequence of recorded user behaviors, where multiple users share a single account, and their behaviours are available in multiple domains.
Existing work on solving \ac{SCSR} mainly relies on mining sequential patterns via RNN-based models, which are not expressive enough to capture the relationships among multiple entities. Moreover, all existing algorithms try to bridge two domains via knowledge transfer in the latent space, and the explicit cross-domain graph structure is unexploited.
In this work, we propose a novel graph-based solution, namely DA-GCN, to address the above challenges. Specifically, we first link users and items in each domain as a graph. Then, we devise a domain-aware graph convolution network to learn user-specific node representations. To fully account for users' domain-specific preferences on items, two novel attention mechanisms are further developed to selectively guide the message passing process. Extensive experiments on two real-world datasets are conducted to demonstrate the superiority of our DA-GCN method.
\end{abstract}
\vspace{-0.5cm}
\section{Introduction}
\ac{CSR} is a recommendation task that aims at recommending the next item via leveraging a user's historical interactions from multiple domains. \ac{CSR} is gaining immense research attention nowadays as users need to sign up for different platforms to access domain-specific services, e.g., music subscription and food delivery.
In this work, we study \ac{CSR} in an emerging yet challenging scenario, \ac{SCSR}, where multiple individual users share a single account and their interaction behaviors are recorded in multiple domains~\cite{DBLP:conf/sigir/Lu0CNZ19}.
We consider the shared-account scenario because it is increasingly common that people share accounts with others in many applications. For example, members of a family tend to share an account for watching movies (e.g., Netflix) and online shopping (e.g., Amazon). Consequently, generating accurate recommendations is much more challenging in this case because of the mixture of diverse users' interests within an interaction sequence~\cite{yin_overcoming_2020,yin2019social,guo2020group}. Furthermore, though in a single-user scenario, data from two domains might collaboratively help uncover her/his preferences, the existence of shared accounts will instead amplify the noise in the interaction data and impede the sequential recommendation accuracy.
% We conduct \ac{SR} in a cross-domain scenario since it is a common phenomenon for users to utilize different platforms to access domain-specific services in their daily life. Intuitively, the user interactions in one domain might be helpful for improving recommendations in another domain. 

Recently, several studies have been focused on recommendation with either shared-account or cross-domain settings, very few of them address \ac{SCSR} that simultaneously considers both aspects. In prior work on shared accounts~\cite{bajaj2016experience}, a common approach is to capture the user relationships under the same account with latent representations, but none of them consider the cross-domain context, and are hence inapplicable to \ac{SCSR}. Despite that cross-domain recommenders ~\cite{zhuang2018cross}, partially fit this scenario, the cross-domain knowledge is implicitly transferred in the latent space, the explicit structural information bridging two domains are largely unexplored. Zhao et al.~\shortcite{chengzhao2019ppgn} proposed a graph-based model as a solution, but their work ignores the important sequential information and relies on explicit user ratings that are not usually available in both domains. However, those cross-platform methods are not designed to cope with entangled user preferences on shared accounts. One prior work on \ac{SCSR} is the $\pi$-net method~\cite{ma2019pi}, which formulates \ac{SCSR} as a parallel sequential recommendation problem that is solved by an information-sharing network. Another related work is the PSJNet method~\cite{ren2019net}, which improves the $\pi$-net using a split-join strategy. However, these RNN-based methods overwhelmingly target on discovering sequential dependencies, and have limited capability for capturing the complex relationships among associated entities (i.e., users and items) in both domains. As a result, this limits the expressiveness of learned user and item representations, and also overlooks the explicit structural information (e.g., item-user-item paths) linking two domains.

To address the above challenges, we propose a novel graph-based solution, namely \ac{DA-GCN}, for \ac{SCSR}. Specifically, to model the complicated interaction relationships, we first construct a \ac{CDS} graph to link different domains, where users and items in each domain are nodes and their associations are edges. Thereafter, to adapt to the scattered user preferences on shared accounts, we suppose there are $H$ latent users under each account and leverage the message-passing strategy in a domain-aware graph convolution network to aggregate the information passed from directly linked neighbors. Then, node embeddings of users and items are learned by passing information from connected neighbor nodes. Though interactions with items are recorded at the account level, items in different domains will have different attraction to different users under the same account. In a similar vein, different neighbor users/items may have varied importance when learning the properties of an item. Hence, we further design two specialized attention mechanisms to discriminatively select relevant information during the message passing. Consequently, we are capable of modeling the multifaceted interactions as well as transferring fine-grained domain knowledge by considering the structure information.

The main contributions of this work are summarized as:
\begin{itemize}
\item We investigate an emerging yet challenging recommendation task, namely \ac{SCSR}. After pointing out the defects of existing RNN-based solutions, we bring a new take on the \ac{SCSR} problem with a graph-based approach.
\item
We firstly construct an innovative \ac{CDS} graph that explicitly links accounts and items from two domains, and build a novel graph model named \ac{DA-GCN} to attentively learn expressive representations for items and account-sharing users for recommendation.
\item 
We conduct extensive experiments on two real-word datasets, and the experimental results demonstrate the superiority of \ac{DA-GCN} compared with several state-of-the-art baselines.
\end{itemize}
\vspace{-0.4cm}
\section{Related Work}
% This section briefly discusses the related works from three aspects: cross-domain recommendation, shared-account recommendation and GCN-based recommendation.
 \subsection{Cross-domain Recommendation}
As \ac{CR} concerns data from multiple domains, it has been proven useful in dealing with cold-start~\cite{abel2013cross} and data sparsity issues~\cite{pan2010transfer}. Existing studies can be categorized into traditional methods and deep learning-based methods. In traditional methods, there are two main ways in dealing with \ac{CR}. One is to aggregate knowledge between two domains~\cite{hu2018conet}. Another is to transfer knowledge from the source domain to the target domain~\cite{tang2012cross}. Deep learning-based methods are well suited to transfer learning, as they can learn high-level abstractions among different domains~\cite{chengzhao2019ppgn,liu2020transfer}. For example, Zhao et al.~\shortcite{chengzhao2019ppgn} and Liu et al.~\shortcite{liu2020transfer} studied \ac{CR} by leveraging the \ac{GCN}-based techniques, but their methods all rely on explicit user ratings, and can not be directly applied to \ac{SCSR}. $\pi$-net~\cite{ma2019pi} and PSJNet~\cite{ren2019net} are two recently proposed methods for \ac{SCSR}, but their methods are all based on RNNs, which are neither expressive enough to capture the multiple associations nor can model the structure information that bridges two domains.
\vspace{-0.2cm}
\subsection{Shared-account Recommendation}
The task of recommending items for a shared-account is to make recommendations by modeling the mixture of user behaviours, which usually performs user identification first, and then makes recommendations~\cite{zhao2016passenger,jiang2018identifying}. For example, Wang et al.~\shortcite{wang2014user} are the first to study user identification in a novel perspective, in which they employ user preference and consumption time to indicate user's identity. Jiang et al.~\shortcite{jiang2018identifying} introduced an unsupervised framework to judge users within a same account and then learned the preferences for each of them. In their method, a heterogeneous graph is exploited to learn the relationship between items and the metadata.
Ma et al.~\shortcite{ma2019pi} and Ren et al.~\shortcite{ren2019net} argued that this task can be treated in an end-to-end fashion, and can also be improved by simultaneously considering the cross-domain information.

\subsection{\ac{GCN}-based Recommendation}
The development of the graph neural networks ~\cite{wang2019ngcf,wang_origin_2019} has attracted a lot of attention to exploring graph-based solutions for recommender systems~\cite{guo2021hierarchical,wanggraph}. E.g., PinSage~\cite{ying2018grasage} combines random walks with multiple graph convolutional layers on the item-item graph for Pinterest image recommendation. NGCF~\cite{wang2019ngcf} exploits high-order proximity by propagating embeddings on the user-item interaction graph to simultaneously update the representations for all users and items in an efficient way by implementing the matrix-form rule. Qiu et al.~\shortcite{qiu_exploiting_2020} exploits the session-based recommendation problem via proposing a novel full graph neural network to the item dependencies between different sessions.
However, none of them can be directly applied to \ac{SCSR} as they either fail to capture the important sequential information~\cite{wang2020intention} or focus on the recommendation problem in a single domain.
\vspace{-0.3cm}
% We argue that this task can be improved by introducing the shared-account unit and attention mechanism with GCN to model the shared-account and cross-domain characteristics (Details are shown in Section \ref{Section 3}). 
\section{Methodologies}
\subsection{ Preliminaries}
Let $ S_A = \{ A_1, A_2, \dots, A_i\}$ and $ S_B = \{ B_1, B_2, \dots,  B_j\} $ be the behaviour sequences from domain A and B respectively, where $A_i \in\mathcal{A}(1\leq i\leq p)$ is the item index in domain $A$; $\mathcal{A}$ is the item set in domain $A$; $B_j \in  \mathcal{B} (1\leq j\leq q)$ is the item index in domain B; and $\mathcal{B}$ is the item set in domain B. $U=\left\{U_1,U_2,...,U_k,...,U_{|U|}\right\}$  is a subset of accounts, where $U_k \in\mathcal{U}(1\leq k\leq n)$ denotes the account index and $\mathcal{U}$ represents the whole account set. Given $S_A$ and $S_B$, the task of \ac{SCSR} is to recommend the next item based on a user's historical behaviours. The recommendation probabilities for all candidate items in domain A and B are:
\begin{align}
P(A_{i+1}|S_A, S_B)\sim f_A(S_A, S_B);
\end{align}
\vspace{-0.5cm}
\begin{align}
P(B_{j+1}|S_B, S_A)\sim f_B(S_B, S_A),
\end{align}
where $P(A_{i+1}|S_A, S_B)$ is the probability of recommending the next item $A_{i+1}$ in domain A, given $S_A$ and $S_B$. $f_A(S_A,S_B)$ is the learned function to estimate $P(A_{i+1}|S_A,S_B)$. Similar definitions are applied to $P(B_{j+1}|S_B, S_A)$ and $f_B(S_B, S_A)$.

\begin{figure}
    \centering
    \includegraphics[width=8.8cm]{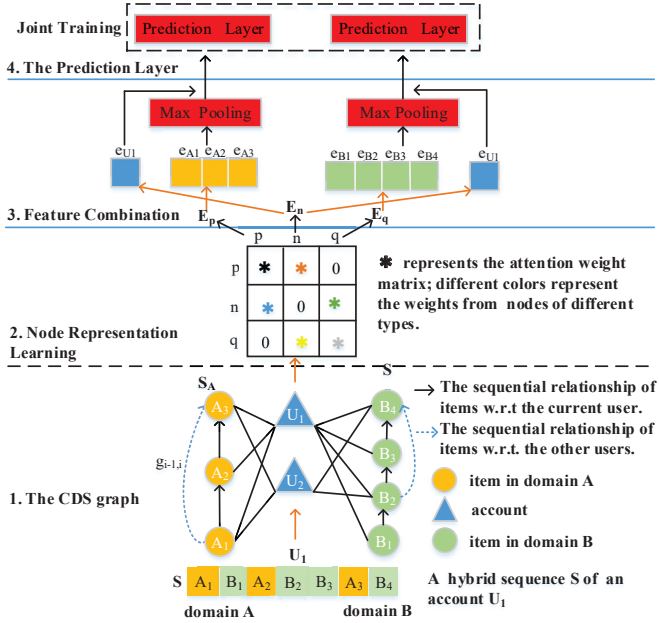}
    \caption{An overview of \ac{DA-GCN}. The directed edge $g_{i-1,i}$ denotes the click of item $A_{i}$ is after item $A_{i-1}$.}
    \label{fig:overview}
\end{figure}

\subsection{An Overview of \ac{DA-GCN}}
The key idea of \ac{DA-GCN} is to learn a graph-based recommendation framework that can model the multiple associations and the structure-aware domain knowledge via recursively aggregating feature information from local graph neighbors in different domains, so as to better capture the diverse user preferences in a shared-account.
Fig.~\ref{fig:overview} presents an overview of \ac{DA-GCN}. Specifically, to model the sophisticated associations among users and items, a \acf{CDS} graph is first constructed. Four types of associations are considered: a) user-item interactions in domain A; b) user-item interactions in domain B; c) sequential transitions among items in domain A; d) sequential transitions among items in domain B. Then, we subsume $H$ latent users under the same account and leverage \ac{GCN} to learn node representation for each latent user, which collects information from the connected neighbors in different domains. In our case, as multiple users interact with different items with a joint account, an interacted item is actually of interest to only some specific account users. Hence, a domain-aware attention mechanism is further proposed to weight the passed messages from different domains. Analogously, we also design another attention scheme on the item side during information aggregation. Thereafter, we generate the sequence-level embedding of each domain via concatenation and pooling operations, which facilitates prediction of the next item. Finally, to mutually maximize the knowledge, a joint training paradigm is leveraged.

\subsection{The Proposed Method}

Motivated by Wang et al.~\shortcite{wang2019kgat} and Zhao et al.~\shortcite{chengzhao2019ppgn}, in this work we employ the message-passing strategy to learn the node representations from the \ac{CDS} graph.
% The details of these learning processes are given in the following.

\subsubsection{Representation Learning with Latent Users}
Since in \ac{SCSR}, an account is usually shared by multiple users, behaviors under the same account are usually generated by different users. Moreover, as users under the same account often have different interests, uniformly treating an account as a virtual user is inappropriate. Because the number and identity of all users under an account are both unknown, we assume there are $H$ latent users ($U_{k,1}, U_{k,2}, ...., U_{k,h}, ..., U_{k,H}$) under each account ($U_k$), and the embedding of user $U_{k,h}$ is denoted by $\bm{e}_{U_{k,h}} \in \mathbb{R}^d$, which is learned by accumulating all the information passed from connected items in domain A and B. Intuitively, by defining latent users under each account, we can learn diversified user representations from their cross-domain interaction sequences to encode multifaceted personalities and preferences.

\textbf{Message Passing.} 
Suppose $A_i\in \mathcal{N}^{U_{k,h}}_{A_i}$ and $B_j\in \mathcal{N}^{U_{k,h}}_{B_j}$ are the item neighbors of $U_{k,h}$ in domain A and B respectively, where $\mathcal{N}^{U_{k,h}}_{A_i}$ is the item neighbor set of $U_{k,h}$ in domain A, and $\mathcal{N}^{U_{k,h}}_{B_j}$ is the item neighbor set in domain B. Specifically, we define the message passed from $A_i\in \mathcal{N}^{U_{k,h}}_{A_i}$ to $U_{k,h}$ as:
\begin{equation}
\bm{m}_{{U_{k,h}} \gets {A_i}}=\gamma_{A_i}^{U_{k,h}} (\bm{W}_1\bm{e}_{A_i}+\bm{W}_2(\bm{e}_{A_i}\odot \bm{e}_{U_{k,h}})),
\end{equation}
where $\bm{m}_{U_{k,h} \gets A_i}$ is the message representation; $\bm{W}_1, \bm{W}_2\in \mathbb{R}^{d^\prime \times d} $ are the trainable weight matrices.
The information interaction between $A_i$ and $U_{k,h}$ is represented by the element-wise product $\bm{e}_{A_i}\odot \bm{e}_{U_{k,h}}$, where $\bm{e}_{A_i} \in \mathbb{R}^d$ and $\bm{e}_{U_{k,h}}$ are the embedding vectors of item $A_i$ and user $U_{k,h}$ respectively.
$\gamma_{A_i}^{U_{k,h}}$ is a learnable parameter that controls how much information can be passed from item $A_i$ to user $U_{k,h}$. We will detail the learning process of this attentive weight in the subsequent section.

Similar to domain A, the message passed from $B_j$ to $U_{k,h}$ in domain B can be defined as:
\begin{equation}
\bm{m}_{{U_{k,h}} \gets {B_j}}=\gamma_{B_j}^{U_{k,h}} (\bm{W}_1\bm{e}_{B_j}+\bm{W}_2(\bm{e}_{B_j}\odot \bm{e}_{U_{k,h}})).
\end{equation}
% \vspace{-0.1cm}
Besides, to retain the information carried by the target user, a self-connection is further added to her/him via:
\begin{equation}
\bm{m}_{{U_{k,h}} \gets {U_{k,h}}}=\gamma_{U_{k,h}}^{U_{k,h}} (\bm{W}_1\bm{e}_{U_{k,h}}),
\end{equation}
where $\gamma_{B_j}^{U_{k,h}}$ and $\gamma_{U_{k,h}}^{U_{k,h}}$ are attentive weights guiding the strength of each message to be passed, which are also computed by an attention mechanism.

\textbf{Domain-aware Attention Mechanism.} On both domains, to fully capture each latent user's distinct preferences over different interacted items, we devise a domain-aware attention mechanism to compute the importance of different item nodes to the target user. The importance of $A_i$ to $U_{k,h}$ in domain A is defined as:
\begin{equation}
s_{A_i}^{U_{k,h}}=f(\bm{e}_{U_{k,h}},\bm{e}_{A_i}),
\end{equation}
where $f(\cdot)$ is a pairwise similarity metric. In this work, cosine similarity is applied. Similarly, the importance of $B_j$ to $U_{k,h}$ in domain B can be defined as:
\begin{equation}
s_{B_j}^{U_{k,h}}=f(\bm{e}_{U_{k,h}},\bm{e}_{B_j}).
\end{equation}

Then, the attentive weight on each item from domain A is obtained by the following normalization operation:
\vspace{-0.3cm}
\begin{equation}
\begin{aligned}
 \gamma_{A_i}^{U_{k,h}} = \rm{exp}(s_{A_i}^{U_{k,h}})/ (\sum_{{A_i}^\prime \in \mathcal{N}_{A_i}^{U_{k,h}}}^{}\rm{exp}(s_{{A_i}^\prime}^{U_{k,h}}) \\
  { + \sum_{B_j^\prime \in \mathcal{N}_{B_j}^{U_{k,h}}}^{}\rm{exp}(s_{{B_j}^\prime}^{U_{k,h}})+ s_{U_{k,h}}^{U_{k,h}}}), \\
 \end{aligned}
\end{equation}
where $s_{U_{k,h}}^{U_{k,h}}$ weighs the self-connection within user $U_{k,h}$. Analogously, by replacing the corresponding user/item embeddings, $\gamma_{B_j}^{U_{k,h}}$ and $\gamma_{U_{k,h}}^{U_{k,h}}$ can be obtained in a similar way.

\textbf{Message Aggregation.} The embedding of user $U_{k,h}$ ($\bm{e}_{U_{k,h}}$) is then updated via:
\vspace{-0.2cm}
\begin{align}
& \bm{e}_{U_{k,h}} = \text{LeakyReLU}(\bm{m}_{{U_{k,h}} \gets {U_{k,h}}} \notag \\
& \qquad 
+ \sum_{A_i\in \mathcal{N}^{U_{k,h}}_{A_i}}{\bm{m}_{U_{k,h} \gets A_i}}+ \sum_{B_j \in \mathcal{N}^{U_{k,h}}_{B_j}}{\bm{m}_{U_{k,h} \gets B_j}}),
\end{align}
where all the messages passed from her/him-self and items in both domains are aggregated, followed by a LeakyReLU activation function.
Eventually, we merge all latent users' representations into the account-level representation $U_k$:
\vspace{-0.2cm}
\begin{align}
    \bm{e}_{U_{k}} = \frac{1}{H}\sum_{h=1}^H \bm{e}_{U_{k,h}}.
    \label{user_add}
\end{align}

\subsubsection{User-specific Item Representation Learning}
For an item, its representation is learned by aggregating the information from two types of nodes, i.e., the connected users and items within the same domain. As the item representations are learned in a similar way in both domains, we take domain A as an example, and the process is also applicable for domain B.

\textbf{Message Passing.} 
Let $A_{i-1}\in \mathcal{N}^{A_i}_{A_{i-1}}$ and $U_{k,h}\in \mathcal{N}^{A_i}_{U_{k,h}}$ be the connected item and user nodes respectively, where $\mathcal{N}^{A_i}_{A_{i-1}}$ is the item neighbor set of $A_i$ in domain A and $\mathcal{N}^{A_i}_{U_{k,h}}$ is the user neighbor set of $A_i$.
Then, the message passed from item $A_{i-1}$ to $A_i$ is formulated as:
\begin{equation}
\bm{m}_{A_i \gets A_{i-1}}=\gamma_{A_{i-1}}^{A_i} (\bm{W}_1{\bm{e}_{A_{i-1}}}+{\bm{W}_2}(\bm{e}_{A_{i-1}}\odot \bm{e}_{A_i})),
\end{equation}
where the parameter $\gamma_{A_{i-1}}^{A_i}$ controls how much information can be passed from $A_{i-1}$ to $A_i$, as we will explain shortly.

The message passed from user $U_{k,h}$ to $A_i$ is defined as:
\begin{equation}
\bm{m}_{A_i \gets U_{k,h}}=\gamma_{U_{k,h}}^{A_i} (\bm{W}_1\bm{e}_{U_{k,h}}+\bm{W}_2(\bm{e}_{U_{k,h}}\odot \bm{e}_{A_i})).
\end{equation}
Similarly, the retained message of $A_i$ is formulated as:
\begin{equation}
\bm{m}_{{A_i} \gets {A_i}}=\gamma_{A_i}^{A_i} (\bm{W}_1\bm{e}_{A_i}),
\end{equation}
where $\gamma_{U_{k,h}}^{A_i}$ and $\gamma_{A_i}^{A_i}$ are both learnable weights.

\textbf{Sequence-aware Attention Mechanism.} Due to varied relevance of linked users/items to the target item, we develop another attention method to assess the importance of the connected users and the items that have sequential relationships to the target item. The importance of item $A_{i-1}$ to $A_i$ is formulated as:
\begin{equation}
s_{A_{i-1}}^{A_i}=f(\bm{e}_{A_{i-1}},\bm{e}_{A_i}).
\end{equation}
The importance of user $U_{k,h}$ on item $A_i$ is defined as:
\vspace{-0.3cm}
\begin{equation}
s_{U_{k,h}}^{A_i}=f(\bm{e}_{A_i},\bm{e}_{U_{k,h}}).
\end{equation}

Then, we normalize $s_{A_{{i-1}}}^{A_i}$ by the following operation:
\vspace{-0.3cm}
\begin{equation}
\begin{aligned}
 \gamma_{A_{i-1}}^{A_i} = \rm{exp}(s_{A_{i-1}}^{A_i})/ (\sum_{{A_{i-1}'} \in \mathcal{N}_{A_{i-1}}^{A_i}}^{}\rm{exp}(s_{{A_{i-1}}^\prime}^{A_i})\\
 { + \sum_{{U_{k,h}}^\prime \in \mathcal{N}_{U_{k,h}}^{A_i}}^{}\rm{exp}(s_{{U_{k,h}}^\prime}^{A_i})+s_{A_i}^{A_i}}),
 \end{aligned}
\end{equation}
where $s_{A_i}^{A_i}$ is the importance of the self-connection within item $A_i$.  $\gamma_{U_{k,h}}^{A_i}$ and $\gamma_{A_i}^{A_i}$ are obtained in the same way.

\textbf{Message Aggregation.} The user-specific item representation is updated by aggregating all messages passed from the neighbor user/item nodes within domain A:
\vspace{-0.3cm}
\begin{align}
& \bm{e}_{A_i} = 
\text{LeakyReLU}(\bm{m}_{{A_i} \gets {A_i}} 
\notag \\
& \qquad
+ \sum_{A_{i-1}\in \mathcal{N}^{A_i}_{A_{i-1}}}{\bm{m}_{A_i \gets A_{i-1}}}+\sum_{U_{k,h} \in \mathcal{N}^{A_i}_{U_{k,h}}}{\bm{m}_{A_i \gets U_{k,h}}}),
\end{align}
For better expression, we use $\bm{g}_{A_i}^h = \bm{e}_{A_i}$ to denote the embedding of item $A_i$ w.r.t the $h$-th user. Then, the final representation of item $A_i$ ($\bm{g}_{A_i}$) can be defined as:
\vspace{-0.3cm}
\begin{align}
    \bm{g}_{A_i} = \frac{1}{H}\sum_{h=1}^H \bm{g}_{A_i}^h.
    \label{user_add}
\end{align}

\subsubsection{Matrix-form Propagation Rule}

To efficiently update the representations for all users and items, we formulate the propagation rule in a layer-wise matrix-form, which is defined as:
\begin{align}
\bm{E}_l=\sigma((\bm{\mathcal L}+\bm{I})\bm{E}_{l-1} \bm{W}_1+\mathcal{L} \bm{E}_{l-1}\odot \bm{E}_{l-1} \bm{W}_2),
\end{align}
where $\bm{I}$ denotes an identity matrix; and $\bm{E}_l\in \mathbb{R}^{(p+n+q)\times d}$ is the representation of all user and item nodes in domain A and B. $\mathcal {L}\in \mathbb{R}^{(p+n+q)\times(p+n+q)}$ is the Laplacian matrix for the \ac{CDS} graph, which is defined as:
\begin{align}
\bm{\mathcal{L}}={
\left[ \begin{array}{ccc}
\bm{Y}_{A_i}{}_{A_{i-1}} & \bm{Y}_{A_i}{}_{U_k} & \bm{0}\\
\bm{Y}_{U_k}{}_{A_i}     & \bm{0}       & \bm{Y}_{U_k}{}_{B_j}\\
\bm{0}      & \bm{Y}_{B_j}{}_{U_k} & \bm{Y}_{B_j}{}_{B_{j-1}},
\end{array} 
\right ]},
\end{align}
where $\bm{Y}_{A_i}{}_{A_{i-1}}\in \mathbb{R}^{p\times p}$ and $\bm{Y}_{A_i}{}_{U_{k}}\in \mathbb{R}^{p\times n}$ are the attention matrices carrying the weights from item and user neighbors to the target item in domain A, respectively;  $\bm{Y}_{U_k}{}_{A_{i}}\in \mathbb{R}^{n\times p}$ represents the weights from item neighbors in domain A to user nodes; $\bm{Y}_{U_k}{}_{B_{j}}\in \mathbb{R}^{n\times q}$ denotes the weights from item neighbors in domain B to the user nodes; $\bm{Y}_{B_j}{}_{U_{k}}\in \mathbb{R}^{q\times n}$ and $\bm{Y}_{B_j}{}_{B_{j-1}}\in \mathbb{R}^{q\times q}$ 
represent the weights from user nodes and item nodes to the target item in domain B, respectively.

\subsubsection{The Prediction Layer}
Through the max pooling operation on the item embeddings within the sequence ($\bm{g}_{A_1}, \bm{g}_{A_2}, ..., \bm{g}_{A_i}$) (or ($\bm{g}_{B_1}, \bm{g}_{B_2}, ..., \bm{g}_{B_j}$)), we can get the sequence-level embedding $\bm{h}'_{S_A}$ (or $\bm{h}'_{S_B}$) for $S_A$ (or $S_B$).
Then, we feed the concatenation of $\bm{h}'_{S_A}$ (or  $\bm{h}'_{S_B}$) and $\bm{e}_{U_k}$ into the following prediction layer:
\vspace{-0.2cm}
\begin{equation}
 \begin{split}
P(A_{i+1}|S_A, S_B) =softmax(\bm{W}_A \cdot [\bm{h}'_{S_A},\bm{e}_{U_k}]^\mathrm{T}+b_A);\\
P(B_{j+1}|S_A, S_B) =softmax(\bm{W}_B \cdot [\bm{h}'_{S_B},\bm{e}_{U_k}]^\mathrm{T}+b_B),\notag
 \end{split}
\end{equation}
\noindent where $\bm{W}_A$, $\bm{W}_B$ are the embedding matrices of all items in domain A and B, respectively; $b_A$, $b_B$ are the bias terms.

Then, we leverage a negative log-likelihood loss function to train \ac{DA-GCN} in each domain:
\begin{align}
L_A(\theta ) = -\frac{1}{|\mathbb{S}|}\sum_{S_A, S_B \in \mathbb{S}}\sum_{A_i \in S_A}\text{log} P(A_{i+1}|S_A,S_B),
\end{align}
\vspace{-0.4cm}
\begin{align}
L_B(\theta ) = -\frac{1}{|\mathbb{S}|}\sum_{S_B, S_A \in \mathbb{S}}\sum_{B_j \in S_B}\text{log}P(B_{j+1}|S_A,S_B),
\end{align}
where $\theta$ represents all the parameters of \ac{DA-GCN} and $\mathbb{S}$ denotes the training sequences in both domain A and B. Both objectives are joined via a multi-task training scheme:
\begin{align}
L(\theta ) = L_A(\theta )+L_B(\theta ).
\end{align}
All the parameters in \ac{DA-GCN} are learned via gradient back-propagation algorithm in an end-to-end fashion.
\vspace{-0.3cm}
\begin{table}
\centering
\small
\begin{tabular}{lll}
\toprule
          & HVIDEO & HAMAZON \\
\hline
        &  E-domain  &  M-domain\\
\#Items    & 3,380   & 67,161      \\
\#Logs     &177,758  & 4,406,924  \\
\hline
        &  V-domain  &  B-domain\\
\#Items    & 16,407   & 126,547     \\
\#Logs     &227,390  & 4,287,240  \\
\hline
\#Overlapped-account    & 13,714   & 13,724    \\
\#Sequences              & 134,349  & 289,160   \\
\#Training-sequences     & 102,182  & 204,477   \\
\#Validation-sequences   & 18,966   & 49,814    \\
\#Test-sequences       & 13,201   & 34,869    \\
\bottomrule
\end{tabular}
\caption{Statistics of the datasets.}
\label{tab:datasets}
\end{table}

\begin{table*}[ht]
  \centering
  \small
    \begin{tabular}{lcccccccc|cccccccc}
    \toprule
    \multicolumn{1}{c}{\multirow{3}[4]{*}{\textbf{Methods}}} & \multicolumn{8}{c|}{\textbf{HVIDEO}} &
    \multicolumn{8}{c}{\textbf{HAMAZON }} \\
    % \midrule
    \cmidrule{2-17}
    & \multicolumn{4}{c}{\textbf{E-domain (\%) }} & \multicolumn{4}{c|}{\textbf{V-domain (\%) }} & \multicolumn{4}{c}{\textbf{M-domain (\%)}} & \multicolumn{4}{c}{\textbf{B-domain (\%)}} \\
    \cmidrule{2-17}
          & \multicolumn{2}{c}{MRR} & \multicolumn{2}{c}{Recall} 
          & \multicolumn{2}{c}{MRR} & \multicolumn{2}{c|}{Recall}
          &
          \multicolumn{2}{c}{MRR} & \multicolumn{2}{c}{Recall}
          & \multicolumn{2}{c}{MRR} & \multicolumn{2}{c}{Recall} 
          \\
          \midrule
          & @5      & @20   & @5    & @20   & @5    & @20   & @5    & @20  & @5     & @20   & @5    & @20   & @5    & @20   & @5    & @20
          \\
    \midrule
    POP   &  1.71 & 2.24     &  2.21     & 6.58  &  2.66         &  3.27    &   5.01       &10.49  & 0.36           & 0.49      & 0.73       & 2.02  &   0.14         &  0.22     & 0.42           & 1.22\\
    Item-KNN &  2.11          &  2.90    &   3.01    &12.11   &    4.43       & 2.93     &   10.48        & 23.93 
    & 1.28           & 1.86      &   2.58      &  9.00     & 3.23          &  4.55     &   6.65      & 20.94 \\
    BPR-MF &    1.34        &  1.64    &   2.74       &   5.83  &    1.21  &  1.36       & 1.88 & 3.38
    &   2.90        &  3.06     &  3.90    & 5.50      & 0.88        &  0.96     & 1.23       & 2.15 \\
    \midrule
    VUI-KNN &   2.03         &  3.48    &   6.36    & 24.27  &    3.44       &2.87      &   16.46        &  34.76
    &  -     &   -    &   -    &  -     &  -     &  -     &   -    &  - \\
    \midrule
    NCF-MLP++ & 3.92    & 5.14  & 7.36   & 20.81  & 16.25   & 17.90  & 26.10   & 43.04
    &   13.68        & 14.21      & 18.44         &  24.31     &   13.67        &  14.05     &  18.14         & 22.08 \\
    Conet & 5.01    & 6.21  & 9.26    & 22.71  & 21.25    & 23.28  & 32.94   & 52.72  
    &  14.64         &  15.12     &  19.25         & 24.46      &15.85            & 16.28      &   20.98          &25.56  \\
    \midrule
    GRU4REC & 12.27   & 13.70  & 16.24    & 32.16  & 78.27    & 78.27  & 80.15    & 83.04  
    &  82.01     & 82.11           & 83.10      & 84.06      &  81.34         &    81.44   &  82.77         &  83.76 \\
    HGRU4REC & 14.47    & 16.11  & 19.79    & 37.52  & 80.37    & 80.62  & 81.92    & 84.43  
    & 83.07          &  83.14     & 84.24            &  84.91     &  82.15      &    82.31   &  83.46        & 84.91 \\
    \midrule
    $\pi$-net & 14.63   & 16.88  & 20.41    & 45.19  & 80.51    & 80.95  & 83.22    & 87.48 
    & 83.91           & 83.95      &  84.91          &    85.33      &  84.93     & 84.93      &  85.33         & 85.38 \\
    PSJNet & 16.63    & 18.46  & 22.12    & 42.20  & 81.97    & 82.32  & 84.32    & 87.75
    & 84.01            &  84.05      &  84.88         &    85.28   & \textbf{85.10}           & \textbf{85.11}     &   85.32        & 85.38 \\
    \hline
    DA-GCN &\textbf{19.24}    & \textbf{21.24}  & \textbf{26.65}    & \textbf{47.78}  & \textbf{83.13}   
    & \textbf{83.42}  & \textbf{85.46}  
    & \textbf{88.30}   
    &  \textbf{84.69}     & \textbf{84.71}  & \textbf{85.13}    & \textbf{85.34}  & 84.81    & 84.81 & \textbf{85.32}    & \textbf{85.38}  \\
    \bottomrule
    \end{tabular}%
  \caption{Experimental results on HVIDEO and HAMAZON. VUI-KNN does not work on this dataset because it needs specific time in a day which is not available on HAMAZON dataset.}
  \label{tab:resutls}%
\end{table*}%

\section{Experiments}
\subsection{Experimental Setup}

\textbf{Research Questions.} We intend to answer the following research questions: 

\textbf{RQ1} How does our proposed \ac{DA-GCN} method perform compared with other state-of-the-art methods?

\textbf{RQ2} Is it helpful to leverage the sequence information? Is it helpful to learn the node representations by incorporating the attention mechanism? 

\textbf{RQ3} Is it helpful to model the shared-account characteristic? How does the hyper-parameter $H$ (the number of users) affect the performance of \ac{DA-GCN}?

\textbf{RQ4} How is the training efficiency and scalability of \ac{DA-GCN} when processing large-scale data?

\textbf{Datasets and Evaluation Protocols.} We evaluate \ac{DA-GCN} on two real-world datasets that are released by~Ren et al.~\shortcite{ren2019net}, i.e., HVIDEO and HAMAZON. HVIDEO is a smart TV dataset that contains the watching logs from two platforms, i.e., the V-domain (the videos of TV series, movies, etc.) and E-domain (the educational videos based on textbooks and instructional videos on sports food, medical, etc.). HAMAZON is the data from two Amazon domains, that is, the movie watching and rating records (M-domain) and the book reading and rating behaviors (B-domain).

We randomly select 75\% of all the sequences as the training data, 15\% as the validation set and the remaining 10\% as the test set. For evaluation, we use the last watched item in each sequence
for each domain as the ground truth and report the results measured by the commonly used metrics MRR@5, Recall@5, MRR@20 and Recall@20. The statistics of the two datasets are shown in Table~\ref{tab:datasets}.

\textbf{Baselines.}
We compare \ac{DA-GCN} with the following baseline methods:
1) Traditional recommendations: POP~\cite{hexiangnan2017ncf}, Item-KNN~\cite{greg2003knn}, and BPR-MF~\cite{hidasi2016srnn}.
2) Shared-account recommendations: VUI-KNN~\cite{wang2014user}.
3) Cross-domain recommendations: NCF-MLP++~\cite{hexiangnan2017ncf}, and Conet~\cite{hu2018conet}.
4) Sequential recommendations: GRU4REC~\cite{hidasi2016srnn}, HGRU4REC~\cite{quadrana2017hrnn}.
5) Shared-account cross-domain sequential recommendations: $\pi$-net~\cite{ma2019pi}, and PSJNet~\cite{ren2019net}.
% , where $\pi$-net and PSJNet are two state-of-the-art methods that are recently proposed for \ac{SCSR}.

\textbf{Implementation Details.}
We implement \ac{DA-GCN} using the Tensorflow framework accelerated by NVidia RTX 2080 Ti GPU. For the initialization strategy, we randomly initialize the model parameters by the Xavier method \cite{glorot2010understanding}. We take Adam as our optimizing algorithm and also apply gradient clipping with range [-5,5] during the training period. For hyperparameters, the learning rate is set to 0.001. The embedding size is set to 100 on HVIDEO and 200 on HAMAZON. To speed up the training process, the batch size are both set to 128. 
The latent user number $H$ is searched in [1-5] with a step size of 1.
All these parameters are tuned on the validation set.
% We  conduct the  one  sample paired t-tests to verify that all improvements are statistically significant with $p<$0.01.
\vspace{-0.2cm}

\subsection{Experimental Results (RQ1)}
Table~\ref{tab:resutls} shows the comparison results of \ac{DA-GCN} over other baselines on HVIDEO and HAMAZON. We have the following observations: 1) Our \ac{DA-GCN} method achieves the best performance on both domains of HVIDEO and outperforms other baselines on HAMAZON in most metrics, which demonstrates the capability of \ac{DA-GCN} in modeling the multiple associations and the user-specific representations, as well as the structure information of the domain knowledge. 2) The methods developed for \ac{SCSR} are significantly better than other baselines, demonstrating the importance of considering the shared-account and cross-domain characteristics simultaneously. 3) \ac{DA-GCN} outperforms the shared account and cross-domain baselines (i.e., VUI-KNN, NCF-MLP++ and Conet) indicating the importance of the sequential information and the effectiveness of our method in modeling users' sequential patterns. 4) From Table~\ref{tab:resutls}, we also observe that our \ac{GCN}-based solution performs better than RNN-based methods ($\pi$-net and PSJNet). This result demonstrates the capability of our \ac{GCN}-based method in modeling the sequential information and the complicated interaction relationships.
\vspace{-0.2cm}

\begin{table}
  \centering
  \small
    \setlength{\tabcolsep}{0.8mm} \begin{tabular}{lcccc|cccc}
    \toprule
    \multicolumn{1}{c}{\multirow{3}[2]{*}{\textbf{Variants}}} & \multicolumn{4}{c|}{\textbf{E-domain (\%)}} &
    \multicolumn{4}{c}{\textbf{V-domain (\%)}} \\
    % \midrule
    \cmidrule{2-9}
          & \multicolumn{2}{c}{MRR} & \multicolumn{2}{c|}{Recall} 
          & \multicolumn{2}{c}{MRR} & \multicolumn{2}{c}{Recall}
          \\
          \midrule
          & @5      & @20   & @5    & @20   & @5    & @20   & @5    & @20 
          \\
    \midrule
    GCN$_{OSA}$  & 19.24       &  20.86          &26.24       & 43.40
    &82.99  & 83.23 &85.23  & 87.66\\
    GCN$_{OA}$  &  19.11         &   20.81    &  26.07         &  43.78  
    & 83.09  & 83.35 & 85.43 &87.88 \\
    GCN$_{OS}$  &   19.21       & 21.19      & 26.37       &   47.04   &83.12  & 83.42 &85.45  & 88.27\\
    \midrule
    \ac{DA-GCN} &  \textbf{19.24}   & \textbf{21.24}  & \textbf{26.65}   & \textbf{47.78} 
    &\textbf{83.13}  &\textbf{83.42}  & \textbf{85.46} & \textbf{88.30}\\
    \bottomrule
    \end{tabular}%
  \caption{Ablations studies on the HVIDEO dataset.}
  \label{tab:ablation}%
\end{table}%

% \section{Experimental Analysis}
\subsection{Ablation Studies of \ac{DA-GCN} (RQ2)}

To further illustrate the impact of different components to \ac{DA-GCN}, we conduct ablation studies on both datasets. Due to space limitation, only results on HVIDEO are presented and similar results are obtained on HAMAZON. The experimental results are shown in Table~\ref{tab:ablation}, where GCN$_{OS}$ is the method that disables the sequential information when constructing the \ac{CDS} graph. GCN$_{OA}$ is the method that disables the attention mechanisms when aggregating the passed messages. GCN$_{OSA}$ is 
a variant of \ac{DA-GCN} that removes both components. From Table~\ref{tab:ablation} we can find that: 1) \ac{DA-GCN} outperforms GCN$_{OSA}$ and GCN$_{OS}$ demonstrating the importance of leveraging the sequential information in modeling users' preferences. 2) \ac{DA-GCN} outperforms GCN$_{OA}$ indicating the importance of weighting the passed messages differently and our model can learn a better node representation via the proposed attention mechanisms.

\begin{figure}
    \centering
    \includegraphics[width=8.5cm]{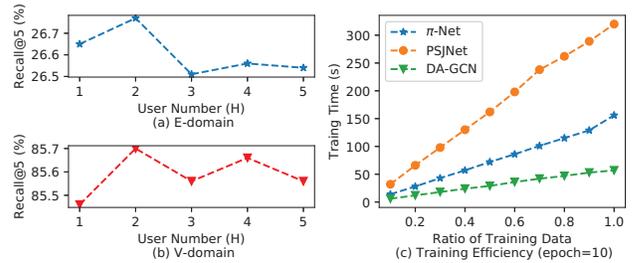}
    \caption{Impact of $H$ and Model Training Efficiency on HVIDEO.}
    \label{fig:training}
\end{figure}

\subsection{Impact of $H$ and Model Training Efficiency}\label{user-size}
\textbf{Impact of $H$ (RQ3).} To explore the importance of modeling the shared-account characteristic, we further conduct experiments on HVIDEO to show the impact of $H$. The experimental results shown in Fig.~\ref{fig:training} demonstrate that viewing an account as a virtual user can not get better results than modeling it as $H$ latent users, which is more in line with reality.

\textbf{Training Efficiency (RQ4).} To investigate the model training efficiency and scalability of our graph-based solution, we further conduct experiments via measuring the time cost for the model training with varying the training ratio of the HVIDEO data in $[0.1, 1.0]$ with step size 0.1. From the experimental results shown in Fig.~\ref{fig:training}, we can find that \ac{DA-GCN} has lower training time cost than $\pi$-net and PSJNet, demonstrating its better scalability to large scale datasets. 
\vspace{-0.2cm}
% \begin{table*}[ht]
%   \centering
%   \footnotesize
%     \begin{tabular}{lcccccccccccc}
%     \toprule
%     \multicolumn{1}{c}{\multirow{3}[2]{*}{\textbf{$H$ values}}} & \multicolumn{6}{c}{\textbf{A-domain recommendation}} & \multicolumn{6}{c}{\textbf{B-domain recommendation}} \\
%     \cmidrule{2-7}\cmidrule{8-13}
%           & \multicolumn{3}{c}{MRR} & \multicolumn{3}{c}{Recall} & \multicolumn{3}{c}{MRR} & \multicolumn{3}{c}{Recall} \\
%           \midrule
%           & @5    & @10   & @20   & @5    & @10   & @20   & @5    & @10   & @20   & @5    & @10   & @20 \\
%     \midrule
%     $H$=1  &       &       &       &       &       &       &       &       &       &       &       &  \\
%     $H$=2  &       &       &       &       &       &       &       &       &       &       &       &  \\
%     $H$=3  &       &       &       &       &       &       &       &       &       &       &       &  \\
    
%     $H$=4  &  \textbf{19.24}  & \textbf{20.39}  & \textbf{21.24}  & \textbf{26.65}  & \textbf{35.45}  & \textbf{47.78}  & \textbf{83.13}   & \textbf{83.31}  & \textbf{83.42}  & \textbf{85.46}  & \textbf{86.82}  & \textbf{88.30}  \\
%     $H$=5  &       &       &       &       &       &       &       &       &       &       &       &  \\
%     \bottomrule
%     \end{tabular}%
%   \caption{Effect of hyperparameter N(\%) on the HVIDEO dataset.}
%   \label{tab:member_number}%
% \end{table*}%

\subsection{Conclusions}
In this work, we propose \ac{DA-GCN} for \ac{SCSR} to fully learn user and item representations from their interactions, as well as the explicit structural information. Specifically, to model the multiple associations among users and items, we first link them in a \ac{CDS} graph. To model the structure information of the transferred knowledge, we then develop a domain-ware \ac{GCN} to learn user-specific node representations, where two attention mechanisms are devised to weight the local neighbors of the target. The experimental results on two real-world datasets demonstrate the superiority of our graph-based solution.

\section*{Acknowledgments}
This work was supported by National Natural Science Foundation of China (Nos. 61602282, 61802236), ARC Discovery Project (No. DP190101985) and ARC Training Centre for Information Resilience (No. IC200100022).

\bibliographystyle{named}
\bibliography{ijcai21}

\end{document}